\RequirePackage{fix-cm}
\documentclass[smallextended]{svjour3}       
\smartqed  
\usepackage{graphicx}
\usepackage{amsmath}
\usepackage{ulem}

 \usepackage{mathptmx}      
\usepackage{graphicx}

\def\art{ART-XC}

\newcommand*{\mysim}{\mathord{\sim}}
\newcommand*{\degr}{$^{\circ}$}

\begin{document}

\title{Calibration of the \art\ mirror modules at MSFC
}

\subtitle{}
\titlerunning{ART-XC mirrors calibration at MSFC}        

\author{R.~Krivonos$^{1}$ \and A.~Tkachenko$^{1}$ \and
  R.~Burenin$^{1}$ \and E.~Filippova$^{1}$ \and I.~Lapshov$^{1}$ \and
  I.~Mereminskiy$^1$ \and S.~Molkov$^{1}$ \and M.~Pavlinsky$^{1}$ \and
  S.~Sazonov$^{1,2}$ \and \fbox{M.~Gubarev}$^{3}$ \and
  J.~Kolodziejczak$^{3}$ \and S.L.~O'Dell$^{3}$ \and D.~Swartz$^{3}$ \and Vyacheslav~E.~Zavlin$^{3}$ \and  B.D.~Ramsey$^{3}$}

\authorrunning{R. Krivonos et al.} 

\dedication{Dedicated to Mikhail Gubarev}

\institute{R. Krivonos 
  \at $^{1}$Space Research Institute of the Russian Academy of
  Sciences,\\ Profsoyuznaya Str. 84/32, 117997 Moscow, Russia;\\
\email{krivonos@iki.rssi.ru} 
           \and
$^{2}$Moscow Institute of Physics and Technology, Institutsky per. 9, 141700 Dolgoprudny, Russia\\
$^{3}$NASA/Marshall Space Flight Center, Huntsville, Alabama 35812, USA\\ 
}

\date{Received: 23 May 2017 / Accepted: 28 Jul 2017}

\maketitle

\begin{abstract}
  The Astronomical R\"ontgen Telescope X-ray Concentrator (\art)
  is a hard X-ray telescope with energy response up to 30
  keV, to be launched on board the Spectrum R\"ontgen Gamma
  (SRG) spacecraft in 2018. \art\ consists of seven identical
  co-aligned mirror modules. Each mirror assembly is coupled with a CdTe
  double-sided strip (DSS) focal-plane detector. Eight X-ray mirror
  modules (seven flight and one spare units) for \art\ were
  developed and fabricated at the Marshall Space Flight Center (MSFC),
  NASA, USA. We present results of testing procedures
  performed with an X-ray beam facility at MSFC to calibrate the point spread function
  (PSF) of the mirror modules. The shape of the PSF was measured with
  a high-resolution CCD camera installed in the focal plane with
  defocusing of 7~mm, as required by the \art\ design. For each
  module, we performed a parametrization of the PSF at various angular
  distances $\Theta$. We used a King function to approximate the radial
  profile of the near on-axis PSF ($\Theta<9$~arcmin) and an ellipse
  fitting procedure to describe the morphology of the far off-axis angular
  response ($9<\Theta<24$~arcmin). We found a good agreement
  between the seven \art\ flight mirror modules at the level of
  10\%. The on-axis angular resolution of the \art\ optics varies between 
  27 and 33~arcsec (half-power diameter), except for the spare module.
\end{abstract}

\section{Introduction}

The Spectrum-R\"ontgen-Gamma (SRG) mission is a Russian-German X-ray
astrophysical observatory that carries two co-aligned X-ray telescope
systems, the German-led extended ROentgen Survey with an Imaging
Telescope Array (eROSITA, \cite{2016SPIE.9905E..1KP}) and the
Russian-led Astronomical R\"ontgen Telescope X-ray Concentrator (\art,
\cite{2016SPIE.9905E..1JP}). Both telescope systems consist of 7
independent mirror modules and focal plane detectors, operating in
$0.2-10$~keV (eROSITA) and $5-30$~keV (\art). SRG is planned to be
launched in 2018 using a Proton-M rocket with a Blok DM-2
upper stage from the Baikonur Cosmodrome into an orbit around L2.

eROSITA, the primary instrument of the SRG mission, will perform a
deep survey of the entire sky in the soft (0.5-2 keV) and medium (2-10
keV) X-ray bands, providing the most sensitive X-ray survey ever
made. \art\ will extend the energy coverage of SRG up to
$\mysim30$~keV, contributing to the eROSITA survey by gathering
complementary information about celestial hard X-ray sources. \art\ will 
perform an all-sky hard X-ray survey almost unaffected by the obscuration 
bias typical for softer X-ray surveys, and provide a census of intrinsically 
strongly absorbed objects such as Compton thick active galactic nuclei 
(AGNs) and heavily obscured Galactic X-ray binary systems. Science
goals of the \art\ telescope include studies of massive nearby galaxy 
clusters; timing and broad band spectroscopy (up to 30 keV) of Galactic objects 
(including X-ray binaries, anomalous pulsars and supernova remnants); 
search for cyclotron line features in X-ray pulsar spectra; exploration of non-thermal 
components in the Galaxy diffuse emission; etc.

Accurate knowledge of an X-ray telescope's response to a point X-ray source is 
essential for its effective use. Here, we report the results of our calibrations of the angular
response of the \art\ mirror modules using high-resolution CCD images
of the telescope's point spread function (PSF) provided by MSFC. We have
performed a parametrization of the PSF at different off-axis distances,
which will be included into the calibration data base (CALDB) of the
\art\ telescope, part of the data reduction software and high-level
scientific analysis.

\section{\art\ Mirror Modules}
\label{sec:art}

The X-ray mirror modules for \art\ were designed and fabricated at
MSFC. Four flight modules were produced under an International
Reimbursable Agreement between NASA and the Space Research Institute
(IKI) in Moscow, Russia, and three flight modules and one spare unit
were fabricated under a Cooperative Agreement between NASA and
IKI. The details of the mirror design and development can be found in
\cite{2012SPIE.8443E..1UG,2013SPIE.8861E..0KG,2014SPIE.9144E..1VG}. Each
module consists of 28 concentrically nested (two-reflection)
grazing-incidence mirrors made of an electroformed nickel-cobalt alloy
and mounted on a single spider through combs glued onto the spider
legs. The spider of each mirror module is fixed to the support plate
on the top of the optical bench. The shells vary in thickness from 250
$\mu$m (inner) to 350 $\mu$m (outer) and range in diameter from about
50 mm to 150 mm. The total length of the primary and secondary mirror
surfaces is 580 mm. The ART-XC shell mirrors are coated with iridium
(Ir) to enhance their high-energy reflectivity up to 30~keV
\cite{2003SPIE.4851..631R,2005ExA....20...85R}. The nominal focal
length of the ART-XC mirror modules is 2700 mm. The actual values for
the seven flight modules are in good mutual agreement (see below).

The \art\ field of view (FOV) is $36$~arcmin in diameter, which is
mainly determined by the working area of the CdTe focal plane detectors
($28.56\times28.56$~mm, \cite{2016SPIE.9905E..1JP}). According to
a raytrace simulation model of the \art\ mirror optics, the
effective-area drop (vignetting) at the FOV edge is about 20\% at
$8$~keV (Tkachenko et al., in preparation).

A number of angular resolution measurements were performed to determine
the optimal angular resolution across the FOV. It was found that 
defocusing a module by intra-focal 7~mm provides more uniform
angular resolution across the FOV compared to the resolution at the
nominal focal distance \cite{2014SPIE.9144E..1VG}. In the following,
we adopt each \art\ module being defocused by 7~mm as a default \art\
design, unless otherwise stated.

The flight-grade mirror modules numbered as ARTM1-5,7,8 were
transported to Sarov (Russia) in 2016 for the final assembly with
the telescope baffle system and focal plane detectors. One spare
module, ARTM6, was delivered to IKI for testing at a 60-meter
X-ray beam facility. The flight model of the \art\ telescope with all
seven mirror modules, after acceptance tests, was transported to
the Lavochkin Association (Khimki, Russia) in December 2016 for 
integration with the SRG spacecraft.

\section{\art\ mirror tests}
\label{sec:msfc}

The \art\ mirror modules were extensively tested with a 104-meter
X-ray beam facility at MSFC \cite{2014SPIE.9144E..1VG}. The distance
from the radiation source to the principal plane of the mirror module,
i.e., the object distance, was 103.3 m. In the thin lens
approximation, the nominal focal length 2700 mm results in an image
distance of 2772 mm. To measure the spatial shape of the PSF at the
best available quality, a high-resolution CCD sensor ANDOR iKon-L 936
with a pixel size of $13.5\times13.5 \mu{\rm m}$ was installed 2765 mm
from the mirror module's principal plane, resulting in a defocus
of 7 mm intra-focal.

The radiation source was a copper anode of the X-ray tube with applied
15 kilovolt voltage. The spectrum of the X-ray source is thus
dominated by copper (Cu) K$\alpha$ and K$\beta$ lines, at 8.04 and 8.9
keV, respectively. Thus, the angular response of the \art\ mirrors was
calibrated at MSFC using an almost monochromatic X-ray beam at energy
$\mysim8$~keV. Note that the testing procedures were done without the
baffle of the telescope, which was installed later, after assembling
the telescope in Sarov.

\begin{figure}
\centerline{\includegraphics[angle=-90,clip,width=\textwidth]{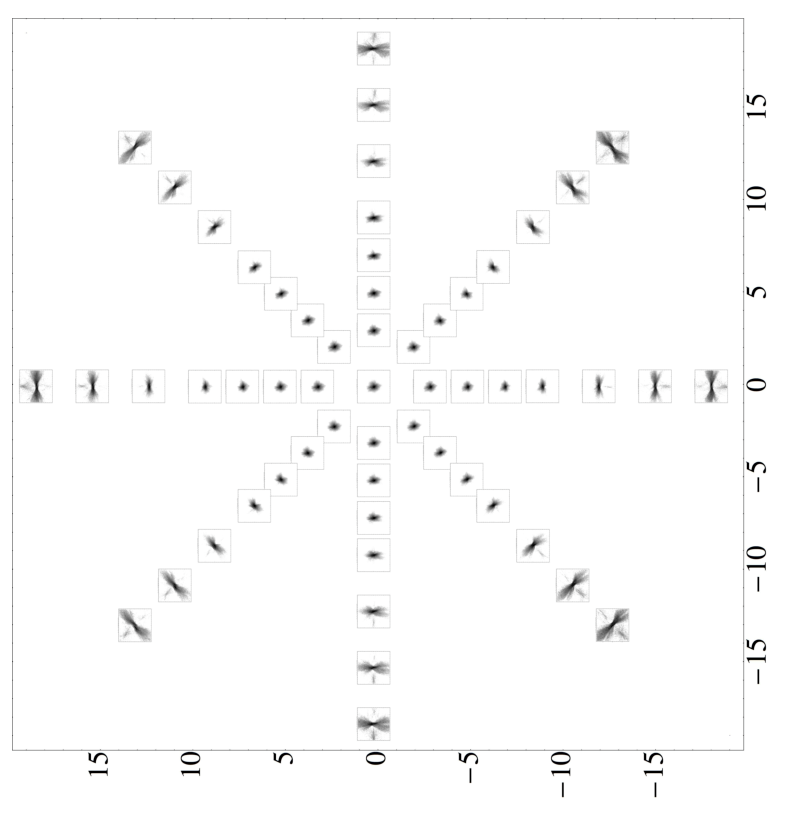}}
\caption{The mosaic illustrates the PSF of ARTM1 in scans at different
  azimutal angles $\Phi$ = 0\degr, 45\degr, 90\degr and 135\degr, and angular
  distances $\Theta=$ 0, 3, 5, 7, 9, 12, 15 and 18 arcmin. PSF images at 1
  arcmin are not shown for clarity.}\label{fig:mosaic}
\end{figure}

\begin{figure*}
\centerline{\includegraphics[width=\textwidth]{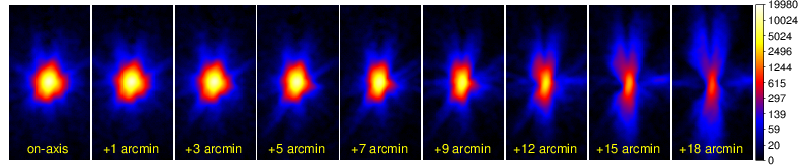}}
\caption[ewewe]{A sample of nine original ARTM1 PSF images at
  different angular distances $\Theta=0-18$~arcmin from the miror's
  optical axis. One image pixel corresponds to $2\times13.5\mu m$ or
  $\mysim$2 arcsec, calculated for a detector distance of 2765 mm
  (7-mm intra-focal defocus). The size of each image box is
  $\mysim2\times4$~mm or $\mysim150\times300$~arcsec. Hereafter, the
  PSF images are background subtracted and smoothed with a 3~pixel width
  tophat filter in {\sc DS9} (SAOImage astronomical imaging and data
  visualization application) for better view. The logarithmic color
  map ranges from 0 to the peak value of the on-axis
  PSF.}\label{fig:artm1:sample}
\end{figure*}

Each \art\ mirror module was tilted with respect to the X-ray beam in
order to obtain PSF images at off-axis angles $\Theta=$ 0, 1, 3,
5, 7, 9, 12, 15 and 18 arcmin and azimutal angles $\Phi=$ 0\degr,
45\degr, 90\degr and 135\degr, as illustrated in
Fig.~\ref{fig:mosaic}. As seen from a sample of original ARTM1 PSF
images (Fig.~\ref{fig:artm1:sample}), the shape of the angular
response remains nearly regular within 9 arcmin from the optical axis.

\begin{figure*}
\centerline{\includegraphics[width=1.0\textwidth]{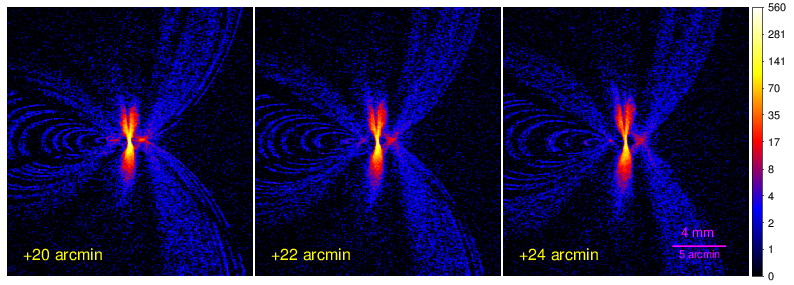}}
\caption{A sample of three original ARTM7 PSF images at extreme
  angular distances $\Theta=$ 20, 22 and 24 arcmin. The size of the
  images is $\mysim18\times20$~mm or $\mysim23\times25$~arcmin.  The
  logarithmic color map ranges from 0 to the peak value of the PSF
  at $\Theta=20$~arcmin.}\label{fig:artm7:extreme}
\end{figure*}

MSFC has also carried out calibrations of ARTM7 angular response at
extreme off-axis distances $\Theta=$ 20, 22 and 24 arcmin
(Fig.~\ref{fig:artm7:extreme}), to simulate a halo of ghost rays
(single-reflection photons), from a bright source located nearly
outside the FOV of the \art\ telescope.

\section{PSF image reduction}
\label{sec:analysis}

One of the main goals of this work is to build a simple analytical
representation of the PSF for each \art\ mirror module at different
off-axis distances. The CCD images of the \art\ PSF provided by MSFC
(Fig.~\ref{fig:mosaic}) contain relevant information about the
construction of the mirror modules, e.g. the shadow of the spider
structure visible at different azimutal directions, misallignment of
the mirror shells, etc. For instance, Fig.~\ref{fig:artm1:sample}
shows that even the on-axis PSF of ARTM1 is characterized by a
non-uniform and slightly assymetric shape, which does not change with
azimutal rotation. We keep the original PSF images in the \art\
calibration data base for possible future use if a detailed knowledge
of the \art\ PSF is needed. Studing the radial profiles of the on-axis
PSF for the different mirror modules, we found that the systematic
deviation between the modules can reach $20-30$\%. The angular size of
the \art\ detector element of $\mysim45$~arcsec is a factor of 1.5
larger than the $\sim30$~arcsec half-power diameter (HPD,
corresponding to half of the focused X-rays) of the optics
\cite{2016SPIE.9905E..1JP}, which allows us to supress detailed
structures of the PSF for simplicity, by averaging them azimuthally at
a given off-axis angle.

\begin{figure*}
\centerline{\includegraphics[clip,width=1.05\textwidth]{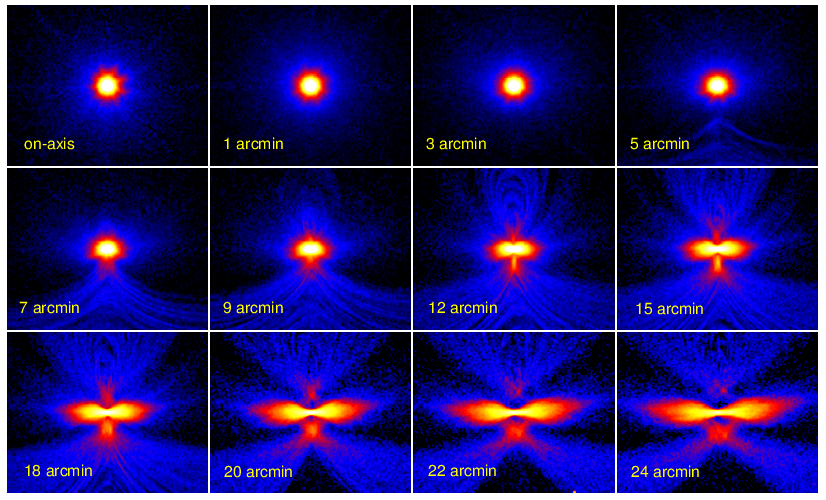}}
\caption{Average PSF images for the ARTM1 mirror module at different
  off-axis distances $\Theta$. The size of each image is
  $\mysim10\times8$~mm or $\mysim13\times10$~arcmin. Note that PSF
  images at $\Theta=$ 20, 22 and 24~arcmin were measured for ARTM7
  only and were ascribed to all mirror modules.}\label{fig:artm1:psf:all}
\end{figure*}

To make an average PSF for each mirror module, we first determined the
centroid position of each raw PSF within the HPD around the PSF
maximum. The relative systematic offset between the centroid positions
did not exceed $135 \mu m$ (or $\mysim10$~arcsec) accross the
FOV. Each PSF image was shifted to the centroid position and rotated
at the corresponding azimutal angle, to align all PSF images in one
direction, with the optical axis position to the north. In this way,
the images were stacked together to produce an averaged PSF image at a
given off-axis distance. Note that we ascribed the stacked PSF images
at angular distances 20, 22 and 24 arcmin measured for ARTM7 only
(Fig.~\ref{fig:artm7:extreme}) to all mirror modules, in order to have
uniform off-axis coverage in the \art\ calibration data
base. Figure~\ref{fig:artm1:psf:all} shows average PSF images for the
ARTM1 mirror module.

\section{Parametrization of the ART-XC Point Spread Function}
\label{sec:model}

In this section, we describe the average PSF shape of each \art\ mirror
module at different off-axis distances. We divide PSF images into two
groups. The first,``near on-axis'', group comprises all off-axis
distances within 9~arcmin from the optical axis. This group is
characterized by a virtually regular shape of the PSF, which can be
represented analytically. The second,``far off-axis', group includes all
PSFs at angular distances larger than 9~arcmin, where off-axis
aberrations are severe. In this case, we apply a simple contoured
parameterization of the PSF, which is nevertheless useful for many analysis
purposes.

\subsection{Near on-axis PSF}
\label{sec:parametrization:onaxis}


\begin{table*}
\noindent
\centering
\caption{Best-fitting model parameters of the decomposition of the \art\ PSF with
  the use of linear combination of two King functions.}\label{tab:king}
\centering
\vspace{1mm}
 \begin{tabular}{|c|c|c|c|c|c|c|c|c|c|c|c|c|c|c|c|c|c|c|}
\hline
Module &\multicolumn{6}{c|}{Off-axis distance $\Theta$ (arcmin)}\\
            & 0 &  1 &  3 & 5 & 7 & 9 \\
\hline

           &\multicolumn{6}{c|}{$N_{\rm core}$ ($\times10^{-3}$~arcsec$^2$)}\\
ARTM1  & $ 1.81\pm 0.14$  & $ 1.85\pm 0.14$  & $ 1.98\pm 0.15$  & $ 1.97\pm 0.15$  & $ 1.76\pm 0.14$  & $ 1.25\pm 0.10$ \\
ARTM2  & $ 1.47\pm 0.11$  & $ 1.53\pm 0.12$  & $ 1.63\pm 0.12$  & $ 1.58\pm 0.12$  & $ 1.40\pm 0.11$  & $ 1.02\pm 0.08$ \\
ARTM3  & $ 1.55\pm 0.12$  & $ 1.59\pm 0.12$  & $ 1.70\pm 0.13$  & $ 1.76\pm 0.14$  & $ 1.64\pm 0.13$  & $ 1.29\pm 0.10$ \\
ARTM4  & $ 1.24\pm 0.09$  & $ 1.26\pm 0.10$  & $ 1.35\pm 0.10$  & $ 1.38\pm 0.11$  & $ 1.29\pm 0.10$  & $ 1.06\pm 0.09$ \\
ARTM5  & $ 1.59\pm 0.12$  & $ 1.62\pm 0.13$  & $ 1.65\pm 0.13$  & $ 1.63\pm 0.13$  & $ 1.41\pm 0.11$  & $ 0.94\pm 0.07$ \\
ARTM6  & $ 0.82\pm 0.06$  & $ 0.85\pm 0.06$  & $ 0.87\pm 0.06$  & $ 0.76\pm 0.05$  & $ 0.74\pm 0.05$  & $ 0.63\pm 0.05$ \\
ARTM7  & $ 1.34\pm 0.10$  & $ 1.39\pm 0.10$  & $ 1.49\pm 0.11$  & $ 1.47\pm 0.11$  & $ 1.30\pm 0.10$  & $ 1.01\pm 0.08$ \\
ARTM8  & $ 1.18\pm 0.09$  & $ 1.24\pm 0.09$  & $ 1.33\pm 0.10$  & $ 1.31\pm 0.10$  & $ 1.16\pm 0.09$  & $ 0.92\pm 0.07$ \\        
&\multicolumn{6}{c|}{$\sigma_{\rm core}$ (arcsec)}\\
ARTM1  & $ 12.8\pm  0.3$  & $ 12.7\pm  0.3$  & $ 12.3\pm  0.3$  & $ 12.0\pm  0.3$  & $ 12.5\pm  0.3$  & $ 14.5\pm  0.4$\\
ARTM2  & $ 14.3\pm  0.4$  & $ 14.0\pm  0.4$  & $ 13.6\pm  0.3$  & $ 13.5\pm  0.3$  & $ 14.0\pm  0.4$  & $ 16.0\pm  0.5$\\
ARTM3  & $ 13.9\pm  0.3$  & $ 13.7\pm  0.3$  & $ 13.3\pm  0.3$  & $ 12.7\pm  0.3$  & $ 12.8\pm  0.3$  & $ 14.1\pm  0.4$\\
ARTM4  & $ 15.5\pm  0.4$  & $ 15.4\pm  0.4$  & $ 14.9\pm  0.4$  & $ 14.5\pm  0.4$  & $ 14.6\pm  0.4$  & $ 15.6\pm  0.5$\\
ARTM5  & $ 13.6\pm  0.4$  & $ 13.4\pm  0.4$  & $ 13.3\pm  0.3$  & $ 13.2\pm  0.4$  & $ 13.9\pm  0.4$  & $ 16.3\pm  0.5$\\
ARTM6  & $ 18.7\pm  0.5$  & $ 18.5\pm  0.4$  & $ 18.1\pm  0.4$  & $ 18.1\pm  0.4$  & $ 18.6\pm  0.5$  & $ 19.6\pm  0.6$\\
ARTM7  & $ 14.9\pm  0.4$  & $ 14.6\pm  0.4$  & $ 14.2\pm  0.4$  & $ 14.0\pm  0.4$  & $ 14.5\pm  0.4$  & $ 16.0\pm  0.5$\\
ARTM8  & $ 15.9\pm  0.4$  & $ 15.5\pm  0.4$  & $ 15.0\pm  0.4$  & $14.8\pm  0.4$  & $ 15.2\pm  0.4$  & $ 16.7\pm  0.5$\\
           &\multicolumn{6}{c|}{$\sigma_{\rm wing}$ (arcsec)}\\
ARTM1  & $  237\pm   27$  & $  252\pm   36$  & $  263\pm   42$  & $  417\pm   90$  & $  325\pm   46$  & $  272\pm   35$ \\
ARTM2  & $  216\pm   35$  & $  215\pm   29$  & $  218\pm   33$  & $  391\pm   80$  & $  319\pm   46$  & $  265\pm   32$ \\
ARTM3  & $  268\pm   48$  & $  263\pm   38$  & $  274\pm   45$  & $  383\pm   72$  & $  338\pm   48$  & $  269\pm   31$ \\
ARTM4  & $  193\pm   30$  & $  196\pm   25$  & $  179\pm   24$  & $  242\pm   36$  & $  239\pm   28$  & $  208\pm   22$ \\
ARTM5  & $  200\pm   27$  & $  198\pm   22$  & $  232\pm   30$  & $  235\pm   29$  & $  232\pm   28$  & $  261\pm   32$ \\
ARTM6  & $  405\pm  112$  & $  394\pm   85$  & $  426\pm  103$  & $  659\pm  239$  & $  405\pm   74$  & $  313\pm   44$ \\
ARTM7  & $  239\pm   49$  & $  238\pm   38$  & $  242\pm   43$  & $  332\pm   66$  & $  276\pm   42$  & $  212\pm   27$ \\
ARTM8  & $  250\pm   49$  & $  245\pm   37$  & $  251\pm   41$  & $  355\pm   67$  & $  299\pm   42$  & $  243\pm   29$ \\
           &\multicolumn{6}{c|}{$f_{\rm wing}$ ($\times10^{-4}$)}\\
ARTM1  & $  1.9\pm  0.4$  & $  1.6\pm  0.4$  & $  1.2\pm  0.3$  & $  1.0\pm  0.2$  & $  2.5\pm  0.5$  & $  6.1\pm  1.1$ \\
ARTM2  & $  2.7\pm  0.9$  & $  2.6\pm  0.7$  & $  1.9\pm  0.6$  & $  1.3\pm  0.3$  & $  3.2\pm  0.6$  & $  8.1\pm  1.5$ \\
ARTM3  & $  1.7\pm  0.5$  & $  1.7\pm  0.4$  & $  1.4\pm  0.4$  & $  1.3\pm  0.3$  & $  2.6\pm  0.5$  & $  6.6\pm  1.1$ \\
ARTM4  & $  4.5\pm  1.4$  & $  4.4\pm  1.2$  & $  4.5\pm  1.3$  & $  3.3\pm  0.8$  & $  6.1\pm  1.2$  & $ 13.0\pm  2.4$ \\
ARTM5  & $  4.2\pm  1.1$  & $  4.1\pm  0.9$  & $  3.2\pm  0.7$  & $  3.9\pm  0.8$  & $  6.1\pm  1.2$  & $ 10.6\pm  1.9$ \\
ARTM6  & $  2.1\pm  0.7$  & $  2.2\pm  0.6$  & $  2.0\pm  0.6$  & $  2.1\pm  0.5$  & $  5.4\pm  1.1$  & $ 12.3\pm  2.3$ \\
ARTM7  & $  2.8\pm  1.0$  & $  2.6\pm  0.8$  & $  2.2\pm  0.7$  & $  2.0\pm  0.5$  & $  4.7\pm  1.0$  & $ 13.1\pm  2.7$ \\
ARTM8  & $  2.6\pm  0.9$  & $  2.7\pm  0.8$  & $  2.2\pm  0.7$  & $  2.0\pm  0.5$  & $  4.6\pm  0.9$  & $ 11.5\pm  2.2$ \\

\hline
\end{tabular}\\
\vspace{3mm}
\end{table*}

\begin{figure}
\centerline{\includegraphics[width=\textwidth]{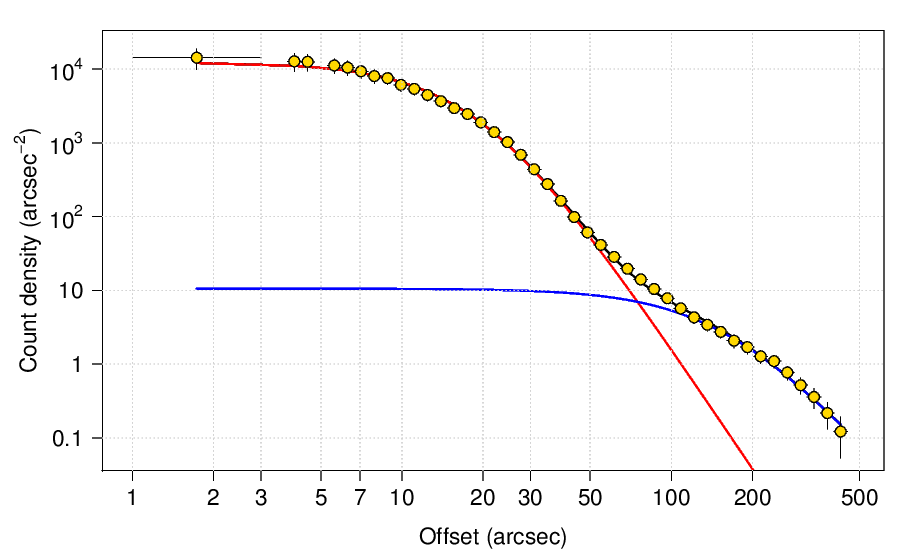}}
\centerline{\includegraphics[width=\textwidth]{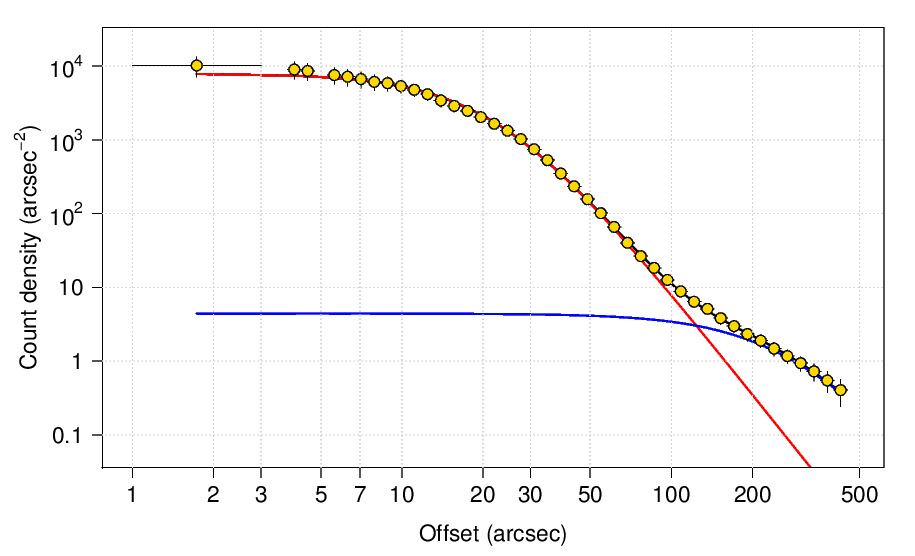}}
\caption{Radial profile of the on-axis PSF measured with the flight ARTM1
  (top panel) and spare ARTM6 (bottom panel) mirror modules in 7~mm
  defocused regime. The King profiles for the PSF core and wing are shown
  with solid red and blue lines, respectively. }\label{fig:artm1:psf}
\end{figure}

Figure~\ref{fig:artm1:psf} shows the radial profile of the on-axis PSF for
the ARTM1 mirror module. The PSF shape of the Wolter-I optics is
characterized by a narrow core and wide wings at offsets larger than 100
arcsec (see e.g. the {\it NuSTAR} optics' calibration
\cite{2015ApJS..220....8M}). Both components can be approximated in
terms of King's function \cite{1962AJ.....67..471K}:
\begin{equation}
K(r,\sigma,\gamma) = \left(1+\frac{r^{2}}{\sigma^{2}}\right)^{-\gamma}
\label{eq:king}
\end{equation}
The parameter $\sigma$ is the characteristic size of the core, 
$\gamma$ determines the weight of the tails, and $r$ is the angular offset
in arcsec. We represent the 2D profile of the \art\ near on-axis PSF as a
linear combination of two King functions:
\begin{equation}
\begin{split}
PSF(r) = &~N_{\rm core} \times \big[  f_{\rm core} K(r,\sigma_{\rm
  core},\gamma_{\rm core})\\ 
&+ f_{\rm wing} K(r,\sigma_{\rm wing},\gamma_{\rm wing})\big],
\end{split}
\label{eq:psfform}
\end{equation}
where $N_{\rm core}$ is an arbitrary overall normalization factor and
$f_{\rm core}+f_{\rm wing}=1$. Due to degeneracy between the $\sigma$ and
$\gamma$ parameters, we fixed the latter at $\gamma=2$ both for the
core and the wing. This also makes possible a direct comparison of the
model parameters amongst the mirror modules.

The stacked PSF images contain a large number of counts per pixel ($>10^4$
in the core), which translates into small statistical errors in the
radial profile. As the fitting procedure weights the data according to
the assigned errors, the data points at $r>100$ arcsec with larger errors
are weighted less in the fit, leading to a loss of sensitivity to the non-central
parts of the PSF. To overcome this difficulty, and to take intrinsic
PSF distortions (Sect.~\ref{sec:analysis}) into account, we added
a fractional systematic error of 30\% to the PSF radial profile As can be seen from Fig.~\ref{fig:artm1:psf}, 
the King-function approximation provides a good fit to the radial profile of the \art\
PSF. The best-fitting values of the free parameters $N_{\rm core}$,
$\sigma_{\rm core}$, $f_{\rm wing}$ and $\sigma_{\rm wing}$ for all
\art\ mirror mudules are listed in Table~\ref{tab:king}. Note that we
have renormalized $N_{\rm core}$ so that
\begin{equation}
\int_{\rm FOV} PSF(r) d\Omega = \int_0^{2\pi} \int_0^{\infty} PSF(r)rdrd\phi = 1.
\end{equation}

\begin{figure}
\includegraphics[width=\textwidth]{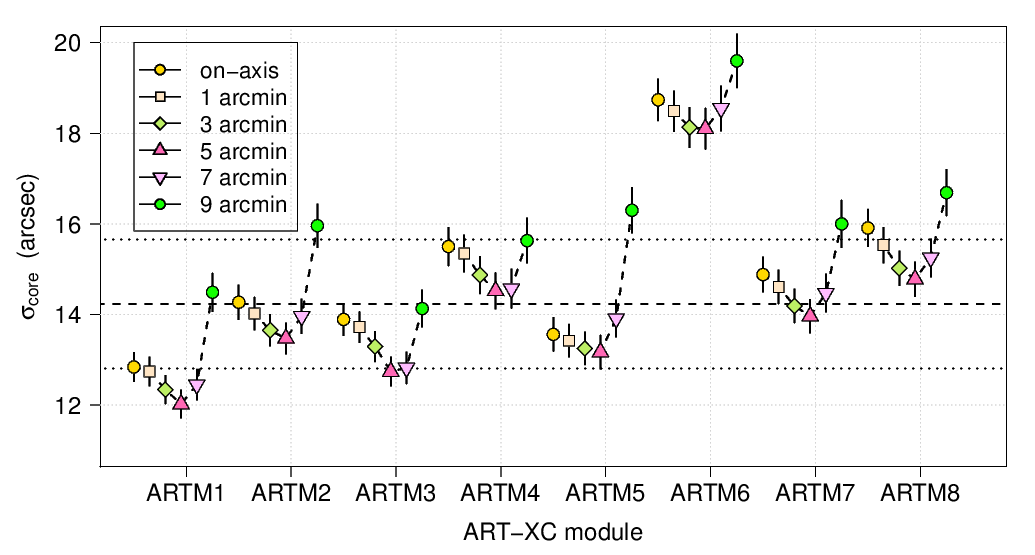}
\includegraphics[width=\textwidth]{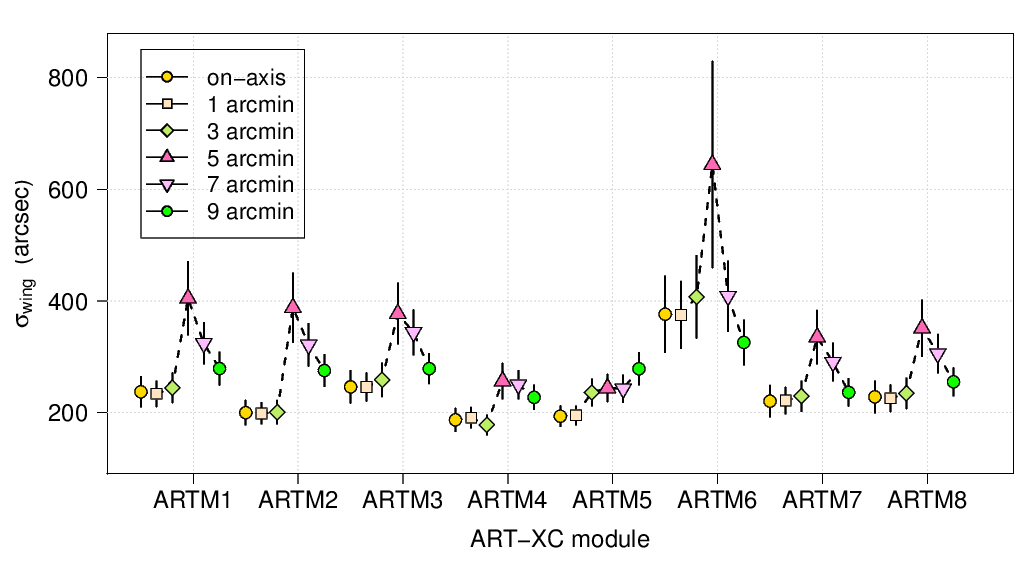}
\caption{Comparison of the PSF core (top panel) and wing (bottom
  panel) characteristic sizes ($\sigma_{\rm core}$ and $\sigma_{\rm
    wing}$ parameters in Eq.~\ref{eq:king}-\ref{eq:psfform},
  respectively) between the \art\ mirror modules for different
  off-axis distances within 9~arcmin. The dashed line in the top panel
  shows the average value of $\sigma_{\rm core}$ 
  ($14.2$~arcsec), derived for the on-axis PSFs and all \art\ mirror
  modules except the spare ARTM6. Dotted lines denote the 10\% interval with
  respect to the average value.}\label{fig:rc_core}
\end{figure}

Figure~\ref{fig:rc_core} compares the best fitting $\sigma_{\rm core}$
parameter between the mirror modules for different off-axis distances
within 9~arcmin.  The PSF core widths for all \art\ flight modules are
consistent with an average $\sigma_{\rm core}$ of 14.2~arcsec to
within 10\%, except for ARTM6, which, as mentioned above, was selected
as a spare unit and will be reserved for future ground tests. As seen
from Fig.~\ref{fig:rc_core}, $\sigma_{\rm core}$ demonstrates a
decrease below $\Theta=$3 arcmin, followed by a rise at larger
off-axis angles. Since the fitting procedure optimizes a linear
combination of two King profiles, $\sigma_{\rm wing}$ shows the
opposite behavior.

\begin{table}
\noindent
\centering
\caption{Total HPD, W90 and the contribution of the wing component to
  the total EEF at $R=\infty$, at a defocus of 7 mm (intra-focal). }\label{tab:hpd}
\centering
\vspace{1mm}
 \begin{tabular}{|c|c|c|c|c|c|c|c|c|c|c|c|c|c|c|c|c|c|c|}
\hline
Module &\multicolumn{6}{c|}{Off-axis distance $\Theta$ (arcmin)}\\
            & 0 &  1 &  3 & 5 & 7 & 9 \\
\hline

           &\multicolumn{6}{c|}{HPD (arcsec)}\\
           ARTM1  &  27.4  &  27.0  &  26.3  &  27.0  &  29.7  &  36.6 \\
           ARTM2  &  30.6  &  29.7  &  28.5  &  30.2  &  33.2  &  40.9 \\
           ARTM3  &  29.7  &  29.3  &  28.1  &  28.5  &  31.0  &  36.6 \\
           ARTM4  &  33.2  &  33.2  &  31.9  &  32.3  &  35.1  &  40.3 \\
           ARTM5  &  29.7  &  29.3  &  29.3  &  30.2  &  33.7  &  43.8 \\
           ARTM6  &  41.5  &  40.9  &  40.3  &  47.6  &  48.3  &  54.7 \\
           ARTM7  &  31.9  &  31.4  &  30.2  &  31.4  &  35.1  &  41.5 \\
           ARTM8  &  33.7  &  33.2  &  31.9  &  33.2  &  36.6  &  43.2 \\

           &\multicolumn{6}{c|}{W90 (arcsec)}\\
           ARTM1  & 120.2  & 116.1  & 108.3  & 286.3  & 449.6  & 516.5 \\
           ARTM2  & 128.8  & 124.5  & 112.1  & 240.7  & 449.6  & 534.8 \\
           ARTM3  & 128.8  & 124.5  & 116.1  & 258.0  & 534.8  & 573.2 \\
           ARTM4  & 143.0  & 143.0  & 133.4  & 182.3  & 365.0  & 465.4 \\
           ARTM5  & 158.7  & 153.3  & 170.1  & 249.2  & 365.0  & 636.1 \\
           ARTM6  & 249.2  & 249.2  & 286.3  &1413.4  & 869.4  & 811.1 \\
           ARTM7  & 143.0  & 138.1  & 128.8  & 258.0  & 449.6  & 481.9 \\
           ARTM8  & 143.0  & 143.0  & 133.4  & 276.5  & 465.4  & 534.8 \\
&\multicolumn{6}{c|}{Wings fraction in EEF at $R=\infty$ (\%)}\\
           ARTM1  &  6.0  &  5.9  &  5.3  & 10.5  & 14.4  & 17.8 \\
           ARTM2  &  6.0  &  5.8  &  4.8  &  9.6  & 14.2  & 18.5 \\
           ARTM3  &  6.1  &  6.0  &  5.6  & 10.1  & 15.4  & 19.6 \\
           ARTM4  &  6.6  &  6.7  &  6.2  &  8.6  & 14.3  & 19.2 \\
           ARTM5  &  8.4  &  8.3  &  8.8  & 11.2  & 14.8  & 21.8 \\
           ARTM6  &  8.8  &  8.9  &  9.5  & 21.2  & 20.5  & 24.3 \\
           ARTM7  &  6.7  &  6.6  &  6.1  & 10.2  & 15.0  & 19.2 \\
           ARTM8  &  6.1  &  6.3  &  5.9  & 10.3  & 15.2  & 19.8 \\
\hline
\end{tabular}\\
\vspace{3mm}
\end{table}

\begin{figure}
\centerline{\includegraphics[width=\textwidth]{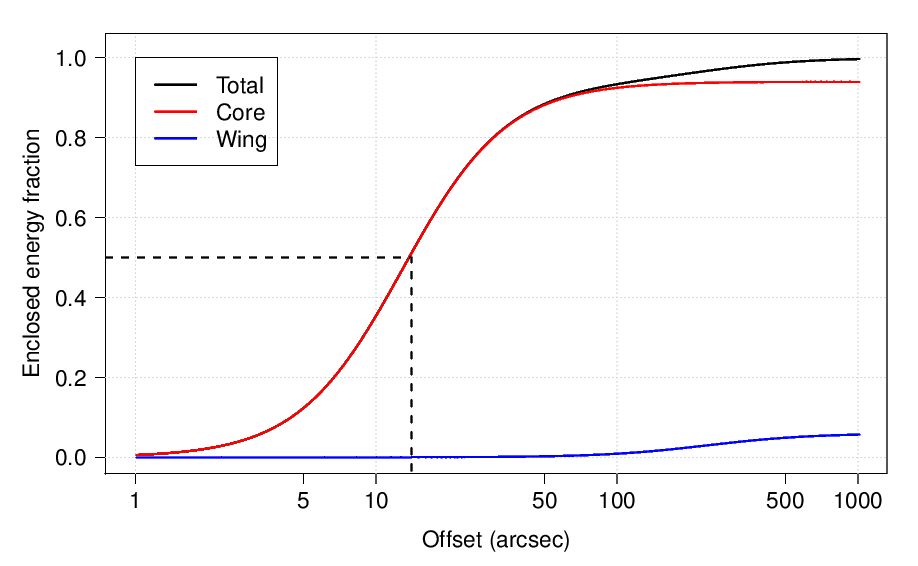}}
\centerline{\includegraphics[width=\textwidth]{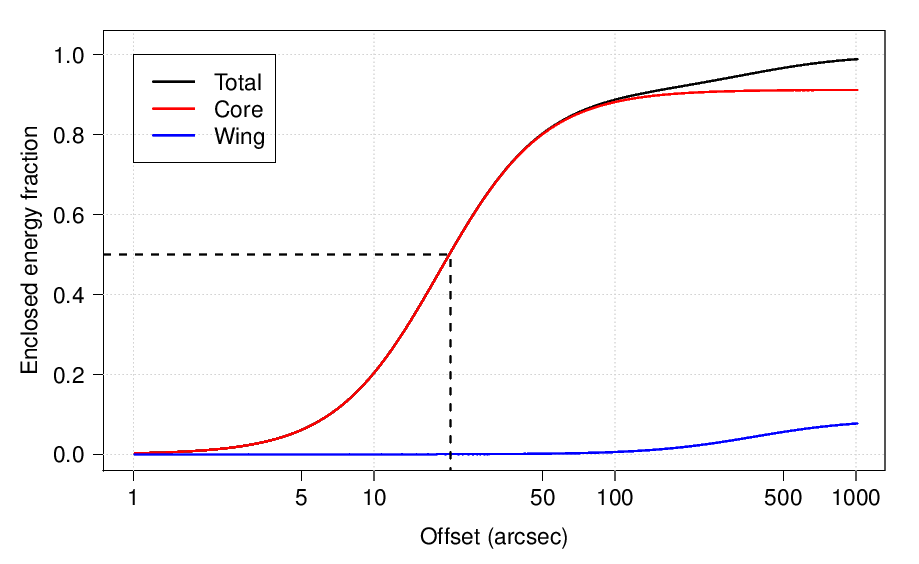}}
\caption{Enclosed energy fraction of the on-axis PSF (top: ARTM1,
  bottom: ARTM6) at 7-mm defocus, derived from the analytic form
  Eq.~\ref{eq:king}-\ref{eq:psfform} for the PSF core (red) and wing
  (blue) components, and total (black). Half of the total EEF,
  corresponding to the HPD value, is shown by the dashed
  line.}\label{fig:eef}
\end{figure}

Using the obtained analytic form (Eq.~\ref{eq:king}-\ref{eq:psfform})
of the PSF, we then calculated the enclosed energy fraction (EEF) as a
function of angular offset $r$ and the corresponding HPD for
each PSF. The typical EEF for the near on-axis \art\ PSF is shown in
Fig.~\ref{fig:eef} (ARTM1 and ARTM6, on-axis). Half of the total
EEF for the King function (Eq.~\ref{eq:king}) with $\gamma=2$ is
reached at an angular offset $r=\sigma$, so that HPD corresponds to
$2\times\sigma$. For the two-component PSF, the presence of broad
wings increases the total HPD by 6\% or more, e.g. for the ARTM1 EEF
shown in Fig.~\ref{fig:eef}, HPD$_{\rm core}=2\times\sigma_{\rm
  core}=25.6\pm0.4$~arcsec, and HPD$_{\rm
  total}=27.4$~arcsec. Table~\ref{tab:hpd} lists the total HPD and W90
(enclosed 90\% energy, diameter) values for each near on-axis \art\
PSF, and the contribution of the wing component to the total EEF
calculated at $r=\infty$.

Finally, we provide a parametrization of the PSF measured in the
nominal focus of the ARTM1 mirror module with on-axis setting, and
tilted at off-axis distances 3 and 7 arcmin
(Table~\ref{tab:king:nominal}).  The total HPD (core+wing) of the
on-axis PSF is $24.2$~arcsec, which is $\mysim12$\% less than the
on-axis HPD in the defocused mode (27.4~arcsec). The difference in HPD
for $\Theta=7$~arcmin and on-axis PSF is 30\%, while in the defocused
mode it is not more than 10\%, which confirms the advantage of the
\art\ defocused mode selected for uniform angular resolution across
the FOV (see Fig.~6 in \cite{2014SPIE.9144E..1VG}).

\begin{table}
\noindent
\centering
\caption{Parametrization of the \art\ PSF with
  the use of linear combination of two King functions
  (Eq.~\ref{eq:psfform}), obtained in the nominal focus mode. The information is available for ARTM1 only.}\label{tab:king:nominal}
\centering
\vspace{1mm}
 \begin{tabular}{|c|c|c|c|c|c|c|c|c|c|c|c|c|c|c|c|c|c|c|}
\hline
Parameter &\multicolumn{3}{c|}{Off-axis distance $\Theta$ (arcmin)}\\
            & 0 &  3 &  7 \\
\hline

$N_{\rm core}$ ($\times10^{-3}$~arcsec$^2$) & $ 2.42\pm 0.20$  & $ 2.43\pm 0.20$  & $ 1.61\pm 0.13$ \\
$\sigma_{\rm core}$ (arcsec) & $ 11.1\pm  0.3$  & $ 11.1\pm  0.3$  & $13.0\pm  0.3$ \\ 
$\sigma_{\rm wing}$ (arcsec) & $  187\pm   16$  & $  201\pm   16$  & $  395\pm   53$ \\
$f_{\rm wing}$ ($\times10^{-4}$) & $  2.7\pm  0.5$  & $  2.0\pm  0.4$  & $  1.9\pm  0.3$ \\
HPD (arcsec)  &  24.2  &  23.8  &  31.0 \\
W90 (arcsec) & 115.5  & 103.4  & 567.5 \\
Wings EEF fraction (\%)  &  7.3  &  6.2  & 14.8 \\
\hline
\end{tabular}\\
\vspace{3mm}
\end{table}

\subsection{Far off-axis PSF}
\label{sec:parametrization:offaxis}

\begin{figure*}
\centerline{\includegraphics[clip,width=\textwidth]{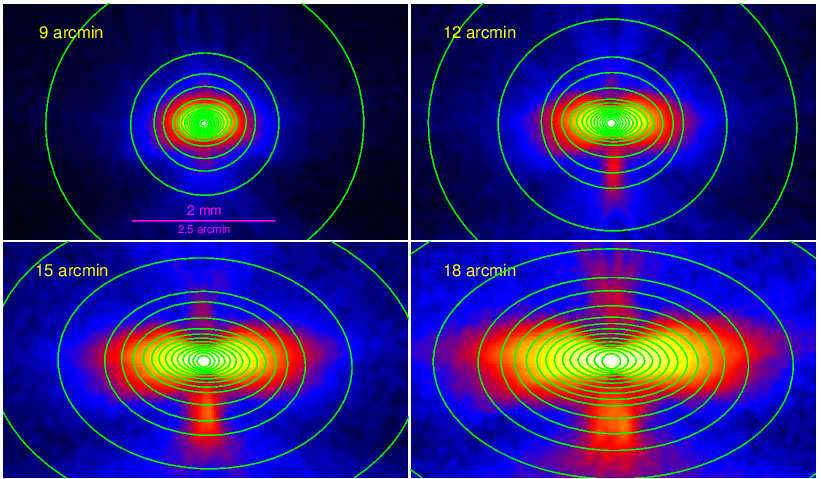}}
\caption{Average ARTM1 PSF images at large off-axis angles of 9, 12, 15
  and 18 arcmin. Green ellipses demonstrate the action of the fitting
  technique, revealing the elongated morphology of the PSF at different
  offset distances.}\label{fig:ellipse}
\end{figure*}

Due to the complex shape of the PSF at large angular distances
($>9$~arcmin), it cannot be easely decomposed into model
components. However, a zero-order description of the PSF can be made in
terms of an encircled energy function, characterizing the PSF with the
fractional energy measured within a circular aperture as a function of
radius. We used this approach in the previous section to compare the \art\
modules with each other for circular-symmetric PSFs at off-axis
distances less than $9$~arcmin.

Similar to other X-ray focusing telescopes, the PSF of the ART-XC optics
becomes elongated with increasing off-axis angle, so that elliptical
apertures are more suitable for representing the underlying PSF. We applied a
simple technique that was developed by \cite{2004SPIE.5165..423A} for fitting 
the ray-traced Chandra PSF at large off-axis angles. The authors introduced an 
enclosed count fraction (ECF) within elliptical regions, analogous to the encircled energy function.
In our study, the difference between ECF and EEF is that the former
defines 100\% counts at a large offset $r\sim15$~arcmin, the
size of the focal plane CCD, whereas the latter calculates the EEF at
$r=\infty$. The ECF and the properties of the elliptical regions
provide a parameterization of the PSF at large off-axis angles, which 
can be used in real observations.

\begin{figure}
\includegraphics[width=\textwidth]{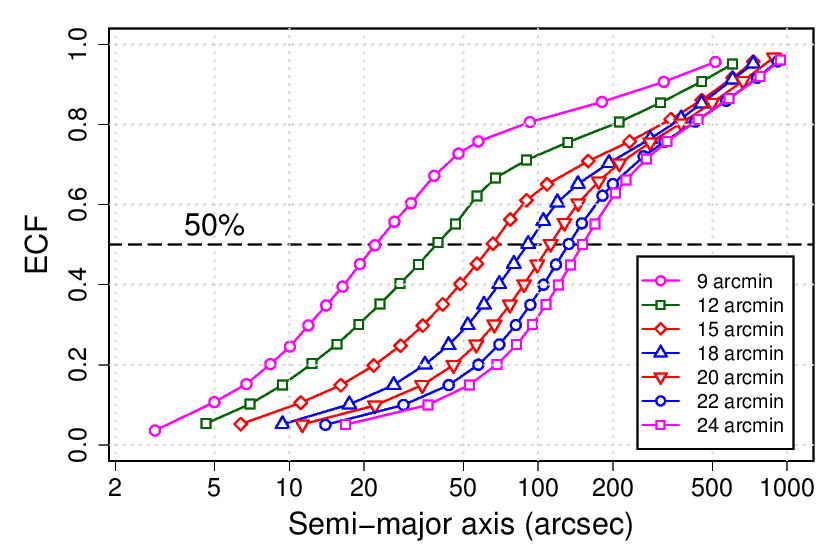}
\caption{ECF within a given ellipse as a function of semi-major axis
  for the ARTM1 PSF at different off-axis distances
  $\Theta=9-24$~arcmin.}\label{fig:ecf:artm1}
\end{figure}


Figure~\ref{fig:ellipse} shows an example of the ellipse fitting technique
applied to the ARTM1 PSF at off-axis distances of 9, 12, 15 and 18
arcmin. The derived ellipses provide a regular and smooth approximation of
the PSF morphology. Note that the ellipse fitting procedure does not
provide ellipse construction at large radii, where the weight of the
data is low (see the method's implementation in \cite{2004SPIE.5165..423A}). 
We thus extended the largest available
ellipse by a linear scaling factor to cover the full range of 
offsets. We constructed ellipses to compose the ECF that are multiples of
5\% within the range 5-95\%. Tables~\ref{tab:ecf} and
\ref{tab:ecf:artm7} list ellipse parameters for ARTM1-5,7,8 at
$\Theta=$ 9, 12, 15 and 18 arcmin, and ARTM7 at $\Theta=$ 20, 22, and
24~arcmin, respectively.  Figure~\ref{fig:ecf:artm1} shows a typical ECF
as a function of ellipse semi-major axis derived for ARTM1.

\begin{figure}
\includegraphics[width=\textwidth]{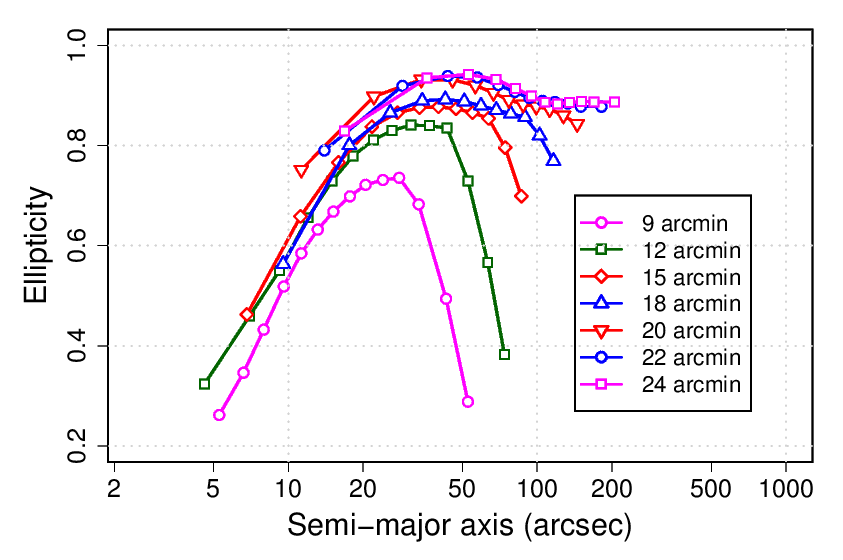}
\caption{Ellipticity as a function of semi-major axis for the ellipses
  used to describe the elongation of the ARTM1 PSF at different
  off-axis distances $\Theta=9-24$~arcmin.}\label{fig:ellipticity}
\end{figure}

The ellipse fitting algorithm reveals different ellipticity of
the PSF at different offset distances $r$ from the center. The shape
of the PSF near the peak, at $r<10$~arcsec, is almost circular, which
implies that the PSF maximum is one-peaked and smooth. At medium offsets
$10<r<70$~arcsec, PSF is strongly elongated, and the overall ellipticity
increases with the off-axis angle. At large offsets $r>100$~arcsec, the
ellipticity decreases, so that the PSF becomes almost circular. 
Figure~\ref{fig:ellipticity} illustrates the ellipticity as a function of semi-major axis 
for the constructed set of ellipses for ARTM1. 

\begin{table*}
\noindent
\centering
\caption{Paramerization of the \art\ PSF in a form of ellipse semi- major
  (a) and  minor (b) axes with a corresponding ECF at off-axis distances 9, 12, 15 and 18 arcmin. The parameters are given only for 10, 50, 75 and 90\% ECF. The list of parameters for the full range 5-95\% ECF is only available in online version of the paper.}\label{tab:ecf}
\centering
\vspace{1mm}
 \begin{tabular}{|r|r|r|r|r|r|r|r|r|r|r|r|r|r|r|r|r|r|r|}
\hline
ECF &\multicolumn{16}{c|}{ART-XC mirror module}\\
 & \multicolumn{2}{c}{ARTM1} &  \multicolumn{2}{c}{ARTM2} 
 & \multicolumn{2}{c}{ARTM3} & \multicolumn{2}{c}{ARTM4} 
 & \multicolumn{2}{c}{ARTM5} & \multicolumn{2}{c}{ARTM6} 
 & \multicolumn{2}{c}{ARTM7} & \multicolumn{2}{c|}{ARTM8} \\
\hline
 & \multicolumn{1}{c|}{a} & \multicolumn{1}{c|}{b} &
\multicolumn{1}{c|}{a} & \multicolumn{1}{c|}{b} &
\multicolumn{1}{c|}{a} & \multicolumn{1}{c|}{b} &
\multicolumn{1}{c|}{a} & \multicolumn{1}{c|}{b} &
\multicolumn{1}{c|}{a} & \multicolumn{1}{c|}{b} &
\multicolumn{1}{c|}{a} & \multicolumn{1}{c|}{b} &
\multicolumn{1}{c|}{a} & \multicolumn{1}{c|}{b} & 
\multicolumn{1}{c|}{a} & \multicolumn{1}{c|}{b} \\
(\%) & \multicolumn{1}{c|}{($''$)}  & \multicolumn{1}{c|}{('')} &
\multicolumn{1}{c|}{($''$)} & \multicolumn{1}{c|}{($''$)} &
\multicolumn{1}{c|}{($''$)} & \multicolumn{1}{c|}{($''$)} &
\multicolumn{1}{c|}{($''$)} & \multicolumn{1}{c|}{($''$)} &
\multicolumn{1}{c|}{($''$)} & \multicolumn{1}{c|}{($''$)} &
\multicolumn{1}{c|}{($''$)}  & \multicolumn{1}{c|}{($''$)} &
\multicolumn{1}{c|}{($''$)} & \multicolumn{1}{c|}{($''$)} &
\multicolumn{1}{c|}{($''$)} & \multicolumn{1}{c|}{($''$)} \\
 \hline
\multicolumn{17}{|c|}{Off-axis 9 arcmin}\\ 
10 &   5.3&  5.1 &   5.0&  4.6 &   5.4&  5.2 &   5.0&  4.6 &   5.0&  4.7 &   7.2&  6.7 &   4.9&  4.6 &   5.1&  4.8 \\
50 &  20.4& 14.2 &  22.1& 14.9 &  20.6& 14.7 &  23.0& 15.0 &  24.6& 16.2 &  31.1& 21.1 &  23.4& 14.7 &  24.1& 15.9 \\
75 &  52.7& 50.4 &  57.5& 54.5 &  57.1& 55.5 &  63.9& 58.7 &  75.8& 64.3 &  98.2& 89.6 &  62.3& 57.3 &  63.5& 60.1 \\
90 & 292.8&280.4 & 319.5&302.9 & 349.1&339.4 & 355.4&326.3 & 383.3&325.1 & 410.2&374.2 & 315.1&289.4 & 321.2&303.6 \\
\multicolumn{17}{|c|}{Off-axis 12 arcmin}\\ 
10 &   7.0&  6.2 &   6.9&  6.2 &   7.3&  6.6 &   6.8&  5.9 &   7.0&  5.8 &   8.7&  8.1 &   6.5&  5.4 &   7.0&  5.9 \\
50 &  36.9& 20.1 &  39.6& 21.1 &  37.1& 20.4 &  40.2& 21.8 &  44.0& 23.4 &  48.6& 27.1 &  40.9& 21.0 &  41.5& 22.5 \\
75 & 118.9&109.9 & 131.4&108.4 & 143.6&126.1 & 153.8&130.0 & 164.8&132.9 & 210.7&165.0 & 140.1&106.9 & 143.5&117.7 \\
90 & 410.5&379.3 & 453.6&374.2 & 450.8&395.8 & 482.6&408.0 & 470.3&379.0 & 546.4&428.0 & 483.7&369.1 & 495.3&406.4 \\
\multicolumn{17}{|c|}{Off-axis 15 arcmin}\\ 
10 &  11.2&  8.4 &  11.1&  8.2 &  11.4&  8.9 &  11.0&  8.0 &  11.0&  7.4 &  11.6& 10.1 &  10.0&  6.8 &  10.1&  7.5 \\
50 &  63.7& 33.1 &  65.8& 34.4 &  64.6& 34.2 &  66.5& 34.8 &  71.2& 36.6 &  79.5& 46.7 &  67.7& 33.8 &  68.2& 35.2 \\
75 & 223.9&160.1 & 232.7&166.7 & 233.9&183.6 & 241.0&180.6 & 255.6&181.4 & 327.5&224.4 & 243.7&161.9 & 248.5&178.6 \\
90 & 580.6&415.2 & 603.5&432.5 & 551.6&433.0 & 568.3&425.8 & 602.6&427.7 & 638.2&437.3 & 632.0&420.1 & 585.9&421.1 \\
\multicolumn{17}{|c|}{Off-axis 18 arcmin}\\ 
10 &  17.6& 10.5 &  17.4&  9.7 &  18.1& 11.3 &  17.9& 10.0 &  18.1&  9.4 &  17.3& 11.6 &  16.3&  8.7 &  15.9&  9.6 \\
50 &  89.3& 46.0 &  91.1& 47.3 &  91.3& 48.4 &  92.5& 49.2 &  98.0& 49.4 & 107.8& 61.1 &  94.3& 45.7 &  93.7& 47.8 \\
75 & 274.0&175.1 & 281.4&180.3 & 285.8&193.5 & 287.3&191.8 & 304.3&191.0 & 356.2&222.8 & 265.7&158.3 & 292.9&187.0 \\
90 & 587.3&375.3 & 603.3&386.6 & 612.6&414.8 & 615.9&411.2 & 652.3&409.5 & 694.1&434.3 & 626.5&373.2 & 627.8&400.7 \\
\hline
\end{tabular}\\
\vspace{3mm}
\end{table*}

\begin{table}
\noindent
\centering
\caption{Paramerization of the PSF for ARTM7 mirror module in a form of ellipse semi- major
  (a) and minor (b) axes with a corresponding ECF at angular distances
  20', 22', and 24'.}\label{tab:ecf:artm7}
\centering
\vspace{1mm}
 \begin{tabular}{|r|r|r|r|r|r|r|r|r|r|r|r|r|r|r|r|r|r|r|}
\hline\hline
 &\multicolumn{6}{c|}{Off-axis distance (arcmin)}\\
 & \multicolumn{2}{c}{20} &  \multicolumn{2}{c}{22} & \multicolumn{2}{c|}{24} \\
\hline
ECF & \multicolumn{1}{c|}{a} & \multicolumn{1}{c|}{b} &
\multicolumn{1}{c|}{a} & \multicolumn{1}{c|}{b} &
\multicolumn{1}{c|}{a} & \multicolumn{1}{c|}{b} \\
(\%) & \multicolumn{1}{c|}{($''$)}  & \multicolumn{1}{c|}{($''$)} &
\multicolumn{1}{c|}{($''$)} & \multicolumn{1}{c|}{($''$)} &
\multicolumn{1}{c|}{($''$)} & \multicolumn{1}{c|}{($''$)} \\
\hline
5  &  11.3&  7.4 &  14.0&  8.6 & 16.8&  9.4 \\
10 &  22.1&  9.7 &  28.8& 11.3 & 36.1& 12.8 \\ 
15 &  34.2& 12.4 & 43.7& 15.0 & 52.9& 17.7 \\
20 &  45.7& 16.6 & 57.5& 20.2 & 68.2& 24.7 \\
25 &  56.4& 22.1 &  69.8& 27.2 & 81.9& 33.2 \\ 
30 &  66.7& 28.3 &   81.3& 34.2 & 95.0& 41.6 \\
35 &  76.9& 34.9 &  92.9& 41.4 &107.4& 49.7 \\
40 &  87.8& 41.3 & 105.0& 48.0 &120.7& 56.6 \\
45 &  99.1& 47.4 & 117.8& 54.4 &135.0& 62.7 \\
50 & 112.1& 54.7 & 132.6& 62.0 &150.8& 69.2 \\
55 & 127.4& 65.0 & 149.7& 71.8 &169.0& 77.9 \\
60 & 144.7& 77.7 & 181.2& 86.8 &204.5& 94.3 \\
65 & 175.0& 94.1 & 199.3& 95.5 &224.9&103.7 \\
70 & 211.8&113.8 & 265.2&127.1 &272.2&125.5 \\
75 & 281.9&151.5 & 320.9&153.8 &329.3&151.9 \\
80 & 375.2&201.6 & 427.2&204.8 & 438.3&202.1 \\
85 & 499.4&268.3 & 568.6&272.5 & 583.4&269.0 \\
90 & 664.7&357.2 & 756.8&362.7 & 776.5&358.1 \\
95 & 884.7&475.4 & 915.7&438.9 &939.6&433.3 \\
\hline
\end{tabular}\\
\vspace{3mm}
\end{table}

\section{Summary and Future Work}

We have analyzed PSF images acquired with a high-resolution CCD camera
at different off-axis angles to build an analytical representation of
the near on-axis \art\ PSF by using a linear combination of King profiles
to describe the core and broad wing of the PSF. We demonstrated that
all seven X-ray optical modules have almost identical angular response, 
with HPD$\sim30$~arcmin, and the deviations between the modules are
within 10\%. The ground testing routines with the X-ray
beam facility at MSFC thus confirm the high angular perfomance of the
\art\ optics. The constrained parameter space enables, through interpolation 
between the reference points, modeling the PSF shape at any off-axis 
distances within 9 arcmin and for each individual \art\ mirror module.

The \art\ PSF at large off-axis distances, characterized by a strong
elongation and shape distortions, was parameterized as a set of
elliptical regions describing the enclosed count fraction (ECF). This
approach allows us to characterize the complex morphology of the PSF
in a simple way, which can be used later on in data reduction.

The constructed PSF model will be compared to testing results obtained  
at the X-ray beam facility at the Space Research Institute (IKI) in Moscow, 
where the spare \art\ mirror module (ARTM6) is being
extensively tested. Although the spare mirror module has the poorest
imaging performance, its characteristics can be readily rescaled for every
flight mirror system using the calibrated parameter space presented
here.

The angular response of the optics was measured at MSFC at
an almost monochromatic X-ray beam of a copper anode at energy
$\mysim8$~keV. The ongoing testing routines at IKI are performed
with the spare CdTe DSS detector, which can detect photons up to
$\mysim150$~keV, well above the useful energy range of the \art\
mirror modules. This allows us to investigate the energy dependence of the
angular response and effective area of the \art\ optics. The results
of the calibration procedures at IKI will be presented elsewhere.

\begin{acknowledgements}
The authors acknowledge support from the Russian Basic
Research Foundation (grant 16-29-13070).
\end{acknowledgements}

\newpage 

\appendix
\setcounter{table}{0}
\renewcommand{\thetable}{A\arabic{table}}
\section*{Appendix: online data}

The following Tables A1-A4 correspond to Table 4 from the main paper
``The calibration of \art\ mirror modules at MSFC'' and contain the
extended paramerization of the ART-XC PSF in a form of ellipse semi-
major (a) and minor (b) axes with a corresponding ECF at off-axis
distances 9, 12, 15 and 18 arcmin. The parameters are given for the
full range 5-95\% ECF.

\begin{table*}[h]
\noindent
\centering
\caption{Paramerization of ART-XC PSF in a form of ellipse semi- major
  (a) and
  minor (b) axes with a corresponding ECF at angular distance 9'.}\label{tab:ecf:09}
\centering
\vspace{1mm}
 \begin{tabular}{|r|r|r|r|r|r|r|r|r|r|r|r|r|r|r|r|r|r|r|}
\hline
 &\multicolumn{16}{c|}{ART-XC mirror module}\\
 & \multicolumn{2}{c}{ARTM1} &  \multicolumn{2}{c}{ARTM2} 
 & \multicolumn{2}{c}{ARTM3} & \multicolumn{2}{c}{ARTM4} 
 & \multicolumn{2}{c}{ARTM5} & \multicolumn{2}{c}{ARTM6} 
 & \multicolumn{2}{c}{ARTM7} & \multicolumn{2}{c|}{ARTM8} \\
\hline
ECF & \multicolumn{1}{c|}{a} & \multicolumn{1}{c|}{b} &
\multicolumn{1}{c|}{a} & \multicolumn{1}{c|}{b} &
\multicolumn{1}{c|}{a} & \multicolumn{1}{c|}{b} &
\multicolumn{1}{c|}{a} & \multicolumn{1}{c|}{b} &
\multicolumn{1}{c|}{a} & \multicolumn{1}{c|}{b} &
\multicolumn{1}{c|}{a} & \multicolumn{1}{c|}{b} &
\multicolumn{1}{c|}{a} & \multicolumn{1}{c|}{b} & 
\multicolumn{1}{c|}{a} & \multicolumn{1}{c|}{b} \\
(\%) & \multicolumn{1}{c|}{('')}  & \multicolumn{1}{c|}{('')} &
\multicolumn{1}{c|}{('')} & \multicolumn{1}{c|}{('')} &
\multicolumn{1}{c|}{('')} & \multicolumn{1}{c|}{('')} &
\multicolumn{1}{c|}{('')} & \multicolumn{1}{c|}{('')} &
\multicolumn{1}{c|}{('')} & \multicolumn{1}{c|}{('')} &
\multicolumn{1}{c|}{('')}  & \multicolumn{1}{c|}{('')} &
\multicolumn{1}{c|}{('')} & \multicolumn{1}{c|}{('')} &
\multicolumn{1}{c|}{('')} & \multicolumn{1}{c|}{('')} \\
 \hline
5  &   2.9&  2.8 &   2.9&  2.6 &   4.1&  3.9 &   3.1&  2.8 &   2.9&  2.7 &   4.9&  4.7 &   2.9&  2.6 &   3.6&  3.3 \\
10 &   5.3&  5.1 &   5.0&  4.6 &   5.4&  5.2 &   5.0&  4.6 &   5.0&  4.7 &   7.2&  6.7 &   4.9&  4.6 &   5.1&  4.8 \\
15 &   6.6&  6.2 &   6.7&  5.9 &   7.0&  6.5 &   6.6&  5.8 &   6.9&  5.9 &   9.6&  8.6 &   6.7&  5.8 &   7.2&  6.2 \\
20 &   8.0&  7.2 &   8.4&  7.0 &   8.8&  7.8 &   8.5&  6.9 &   8.7&  6.8 &  11.5& 10.1 &   8.4&  6.9 &   9.0&  7.5 \\
25 &   9.6&  8.2 &  10.1&  8.1 &  10.1&  8.8 &  10.2&  7.9 &  10.8&  8.2 &  13.7& 11.5 &  10.6&  8.0 &  11.0&  8.7 \\
30 &  11.3&  9.2 &  11.9&  9.2 &  11.7&  9.9 &  12.2&  9.1 &  12.9&  9.3 &  16.3& 12.8 &  12.7&  9.0 &  13.2&  9.9 \\
35 &  13.1& 10.2 &  14.1& 10.4 &  13.5& 11.0 &  14.5& 10.3 &  15.2& 10.6 &  19.2& 14.3 &  14.9& 10.0 &  15.5& 11.1 \\
40 &  15.2& 11.3 &  16.4& 11.6 &  15.7& 12.2 &  16.9& 11.4 &  17.7& 11.9 &  22.4& 16.0 &  17.5& 11.3 &  18.0& 12.4 \\
45 &  17.7& 12.7 &  19.2& 13.2 &  18.1& 13.4 &  19.9& 13.2 &  21.0& 13.9 &  26.2& 18.0 &  20.4& 13.0 &  21.0& 14.1 \\
50 &  20.4& 14.2 &  22.1& 14.9 &  20.6& 14.7 &  23.0& 15.0 &  24.6& 16.2 &  31.1& 21.1 &  23.4& 14.7 &  24.1& 15.9 \\
55 &  24.0& 16.4 &  26.4& 17.6 &  24.2& 16.8 &  27.6& 18.3 &  29.6& 19.5 &  36.6& 24.8 &  27.8& 18.0 &  28.7& 19.2 \\
60 &  27.9& 18.9 &  30.8& 20.5 &  28.4& 19.5 &  32.3& 21.6 &  34.5& 22.9 &  44.8& 33.4 &  32.3& 21.3 &  33.3& 22.6 \\
65 &  33.4& 24.4 &  38.2& 29.0 &  35.4& 27.0 &  41.6& 32.1 &  45.5& 35.2 &  55.9& 47.3 &  40.7& 30.7 &  42.2& 33.1 \\
70 &  43.0& 37.4 &  47.8& 41.7 &  46.2& 41.3 &  52.8& 45.4 &  57.0& 48.3 &  67.1& 61.2 &  51.5& 44.0 &  52.9& 46.6 \\
75 &  52.7& 50.4 &  57.5& 54.5 &  57.1& 55.5 &  63.9& 58.7 &  75.8& 64.3 &  98.2& 89.6 &  62.3& 57.3 &  63.5& 60.1 \\
80 &  77.1& 73.8 &  92.5& 87.7 & 101.1& 98.3 & 102.9& 94.5 & 134.4&114.0 & 174.0&158.7 & 100.4& 92.2 & 102.3& 96.7 \\
85 & 165.3&158.3 & 180.3&171.0 & 197.0&191.6 & 200.6&184.2 & 238.0&201.9 & 280.2&255.6 & 177.9&163.4 & 181.3&171.4 \\
90 & 292.8&280.4 & 319.5&302.9 & 349.1&339.4 & 355.4&326.3 & 383.3&325.1 & 410.2&374.2 & 315.1&289.4 & 321.2&303.6 \\
95 & 518.8&496.7 & 514.5&487.8 & 511.1&496.9 & 520.3&477.8 & 561.2&476.0 & 546.0&498.0 & 507.5&466.2 & 517.2&488.9 \\
\hline
\end{tabular}\\
\vspace{3mm}
\end{table*}

\begin{table*}
\noindent
\centering
\caption{Paramerization of ART-XC PSF in a form of ellipse semi- major
  (a) and
  minor (b) axes with a corresponding ECF at angular distance 12'.}\label{tab:ecf:12}
\centering
\vspace{1mm}
 \begin{tabular}{|r|r|r|r|r|r|r|r|r|r|r|r|r|r|r|r|r|r|r|}
\hline
 &\multicolumn{16}{c|}{ART-XC mirror module}\\
 & \multicolumn{2}{c}{ARTM1} &  \multicolumn{2}{c}{ARTM2} 
 & \multicolumn{2}{c}{ARTM3} & \multicolumn{2}{c}{ARTM4} 
 & \multicolumn{2}{c}{ARTM5} & \multicolumn{2}{c}{ARTM6} 
 & \multicolumn{2}{c}{ARTM7} & \multicolumn{2}{c|}{ARTM8} \\
\hline
ECF & \multicolumn{1}{c|}{a} & \multicolumn{1}{c|}{b} &
\multicolumn{1}{c|}{a} & \multicolumn{1}{c|}{b} &
\multicolumn{1}{c|}{a} & \multicolumn{1}{c|}{b} &
\multicolumn{1}{c|}{a} & \multicolumn{1}{c|}{b} &
\multicolumn{1}{c|}{a} & \multicolumn{1}{c|}{b} &
\multicolumn{1}{c|}{a} & \multicolumn{1}{c|}{b} &
\multicolumn{1}{c|}{a} & \multicolumn{1}{c|}{b} & 
\multicolumn{1}{c|}{a} & \multicolumn{1}{c|}{b} \\
(\%) & \multicolumn{1}{c|}{('')}  & \multicolumn{1}{c|}{('')} &
\multicolumn{1}{c|}{('')} & \multicolumn{1}{c|}{('')} &
\multicolumn{1}{c|}{('')} & \multicolumn{1}{c|}{('')} &
\multicolumn{1}{c|}{('')} & \multicolumn{1}{c|}{('')} &
\multicolumn{1}{c|}{('')} & \multicolumn{1}{c|}{('')} &
\multicolumn{1}{c|}{('')}  & \multicolumn{1}{c|}{('')} &
\multicolumn{1}{c|}{('')} & \multicolumn{1}{c|}{('')} &
\multicolumn{1}{c|}{('')} & \multicolumn{1}{c|}{('')} \\
 \hline
5  &   4.6&  4.4 &   4.6&  4.3 &   4.8&  4.5 &   4.4&  4.0 &   4.5&  4.0 &   5.9&  5.6 &   4.4&  3.8 &   4.6&  4.0 \\
10 &   7.0&  6.2 &   6.9&  6.2 &   7.3&  6.6 &   6.8&  5.9 &   7.0&  5.8 &   8.7&  8.1 &   6.5&  5.4 &   7.0&  5.9 \\
15 &   9.2&  7.7 &   9.4&  7.8 &   9.7&  8.4 &   9.2&  7.1 &  10.2&  7.2 &  11.7& 10.3 &   9.4&  6.8 &   9.9&  7.5 \\
20 &  12.0&  9.1 &  12.4&  8.9 &  12.0&  9.6 &  12.2&  8.2 &  13.2&  8.6 &  14.7& 11.7 &  12.4&  7.8 &  12.8&  8.7 \\
25 &  15.0& 10.3 &  15.5& 10.0 &  15.0& 10.9 &  15.4&  9.5 &  17.0&  9.9 &  18.3& 13.1 &  15.5&  8.8 &  16.2&  9.8 \\
30 &  18.2& 11.4 &  19.0& 11.1 &  18.2& 12.1 &  19.2& 10.9 &  21.0& 11.5 &  22.6& 14.3 &  19.4& 10.1 &  20.0& 11.2 \\
35 &  22.0& 12.8 &  23.1& 12.7 &  21.9& 13.4 &  23.3& 12.5 &  25.6& 13.4 &  27.3& 16.0 &  23.6& 11.9 &  24.2& 12.9 \\
40 &  26.1& 14.6 &  27.9& 14.8 &  26.1& 15.0 &  28.1& 14.6 &  30.8& 15.8 &  33.2& 18.6 &  28.3& 14.2 &  28.9& 15.2 \\
45 &  31.1& 16.8 &  33.1& 17.3 &  30.9& 17.0 &  33.3& 17.2 &  36.7& 18.7 &  40.3& 22.2 &  33.9& 16.9 &  34.3& 17.8 \\
50 &  36.9& 20.1 &  39.6& 21.1 &  37.1& 20.4 &  40.2& 21.8 &  44.0& 23.4 &  48.6& 27.1 &  40.9& 21.0 &  41.5& 22.5 \\
55 &  43.4& 23.9 &  46.5& 25.1 &  43.7& 24.3 &  47.1& 26.5 &  51.8& 28.9 &  60.3& 40.0 &  48.1& 25.4 &  48.8& 27.3 \\
60 &  52.7& 36.0 &  56.6& 39.4 &  55.1& 40.8 &  59.4& 43.6 &  64.3& 45.4 &  73.8& 57.8 &  59.5& 39.1 &  61.2& 43.9 \\
65 &  63.3& 52.1 &  67.4& 55.6 &  67.0& 58.8 &  71.7& 60.7 &  76.9& 62.0 &  98.3& 77.0 &  71.9& 54.9 &  73.6& 60.4 \\
70 &  73.8& 68.2 &  89.7& 74.0 &  89.2& 78.3 &  95.5& 80.7 & 102.3& 82.5 & 143.9&112.7 &  95.7& 73.0 &  98.0& 80.4 \\
75 & 118.9&109.9 & 131.4&108.4 & 143.6&126.1 & 153.8&130.0 & 164.8&132.9 & 210.7&165.0 & 140.1&106.9 & 143.5&117.7 \\
80 & 191.5&176.9 & 211.6&174.6 & 231.3&203.1 & 225.2&190.4 & 241.3&194.5 & 308.4&241.6 & 225.6&172.2 & 231.1&189.6 \\
85 & 280.4&259.1 & 309.8&255.6 & 338.7&297.4 & 329.7&278.7 & 353.3&284.8 & 410.5&321.5 & 330.3&252.1 & 338.3&277.6 \\
90 & 410.5&379.3 & 453.6&374.2 & 450.8&395.8 & 482.6&408.0 & 470.3&379.0 & 546.4&428.0 & 483.7&369.1 & 495.3&406.4 \\
95 & 546.4&504.8 & 603.7&498.1 & 600.0&526.8 & 642.4&543.1 & 625.9&504.5 & 661.2&517.8 & 643.8&491.3 & 659.3&540.9 \\
\hline
\end{tabular}\\
\vspace{3mm}
\end{table*}

\begin{table*}
\noindent
\centering
\caption{Paramerization of ART-XC PSF in a form of ellipse semi- major
  (a) and
  minor (b) axes with a corresponding ECF at angular distance 15'.}\label{tab:ecf:15}
\centering
\vspace{1mm}
 \begin{tabular}{|r|r|r|r|r|r|r|r|r|r|r|r|r|r|r|r|r|r|r|}
\hline
 &\multicolumn{16}{c|}{ART-XC mirror module}\\
 & \multicolumn{2}{c}{ARTM1} &  \multicolumn{2}{c}{ARTM2} 
 & \multicolumn{2}{c}{ARTM3} & \multicolumn{2}{c}{ARTM4} 
 & \multicolumn{2}{c}{ARTM5} & \multicolumn{2}{c}{ARTM6} 
 & \multicolumn{2}{c}{ARTM7} & \multicolumn{2}{c|}{ARTM8} \\
\hline
ECF & \multicolumn{1}{c|}{a} & \multicolumn{1}{c|}{b} &
\multicolumn{1}{c|}{a} & \multicolumn{1}{c|}{b} &
\multicolumn{1}{c|}{a} & \multicolumn{1}{c|}{b} &
\multicolumn{1}{c|}{a} & \multicolumn{1}{c|}{b} &
\multicolumn{1}{c|}{a} & \multicolumn{1}{c|}{b} &
\multicolumn{1}{c|}{a} & \multicolumn{1}{c|}{b} &
\multicolumn{1}{c|}{a} & \multicolumn{1}{c|}{b} & 
\multicolumn{1}{c|}{a} & \multicolumn{1}{c|}{b} \\
(\%) & \multicolumn{1}{c|}{('')}  & \multicolumn{1}{c|}{('')} &
\multicolumn{1}{c|}{('')} & \multicolumn{1}{c|}{('')} &
\multicolumn{1}{c|}{('')} & \multicolumn{1}{c|}{('')} &
\multicolumn{1}{c|}{('')} & \multicolumn{1}{c|}{('')} &
\multicolumn{1}{c|}{('')} & \multicolumn{1}{c|}{('')} &
\multicolumn{1}{c|}{('')}  & \multicolumn{1}{c|}{('')} &
\multicolumn{1}{c|}{('')} & \multicolumn{1}{c|}{('')} &
\multicolumn{1}{c|}{('')} & \multicolumn{1}{c|}{('')} \\
 \hline
5  &   6.8&  6.0 &   6.4&  5.7 &   6.8&  6.2 &   6.5&  5.6 &   5.9&  5.1 &   7.2&  7.0 &   5.6&  4.8 &   6.2&  5.4 \\
10 &  11.2&  8.4 &  11.1&  8.2 &  11.4&  8.9 &  11.0&  8.0 &  11.0&  7.4 &  11.6& 10.1 &  10.0&  6.8 &  10.1&  7.5 \\
15 &  15.9& 10.2 &  16.1&  9.8 &  16.0& 11.0 &  16.0&  9.5 &  16.6&  9.4 &  16.8& 11.8 &  15.1&  8.0 &  15.1&  9.2 \\
20 &  21.7& 11.8 &  21.8& 11.3 &  21.8& 12.4 &  22.1& 11.0 &  23.1& 11.1 &  22.8& 13.5 &  20.9&  9.6 &  20.9& 10.9 \\
25 &  27.5& 13.8 &  28.0& 13.4 &  27.7& 14.2 &  28.2& 13.2 &  30.1& 13.7 &  30.1& 15.6 &  27.5& 11.7 &  27.0& 12.8 \\
30 &  33.7& 16.3 &  34.4& 16.1 &  34.0& 16.4 &  34.8& 15.9 &  37.2& 16.8 &  37.6& 18.6 &  34.1& 14.6 &  33.9& 15.5 \\
35 &  40.1& 19.2 &  41.3& 19.5 &  40.5& 19.4 &  41.5& 19.2 &  44.6& 20.4 &  45.9& 22.4 &  41.6& 18.1 &  41.1& 19.0 \\
40 &  47.2& 23.0 &  48.6& 23.6 &  47.6& 23.1 &  48.7& 23.2 &  52.6& 24.8 &  55.0& 27.2 &  49.3& 22.4 &  49.2& 23.3 \\
45 &  55.0& 27.5 &  56.7& 28.6 &  55.7& 28.2 &  57.2& 28.5 &  61.6& 30.3 &  66.2& 34.0 &  57.9& 27.5 &  58.4& 28.9 \\
50 &  63.7& 33.1 &  65.8& 34.4 &  64.6& 34.2 &  66.5& 34.8 &  71.2& 36.6 &  79.5& 46.7 &  67.7& 33.8 &  68.2& 35.2 \\
55 &  74.3& 45.0 &  77.2& 47.6 &  77.1& 51.7 &  79.5& 51.5 &  84.5& 52.5 &  94.9& 65.0 &  80.1& 46.3 &  81.8& 51.5 \\
60 &  86.3& 61.7 &  89.7& 64.3 &  90.2& 70.8 &  92.9& 69.6 &  98.5& 69.9 & 126.3& 86.5 &  93.9& 62.4 &  95.8& 68.9 \\
65 & 104.4& 74.7 & 108.5& 77.8 & 120.1& 94.2 & 123.7& 92.7 & 131.1& 93.1 & 168.0&115.2 & 113.7& 75.6 & 127.5& 91.6 \\
70 & 152.9&109.3 & 158.9&113.9 & 175.8&138.0 & 181.1&135.7 & 174.6&123.9 & 246.0&168.6 & 166.4&110.6 & 169.7&122.0 \\
75 & 223.9&160.1 & 232.7&166.7 & 233.9&183.6 & 241.0&180.6 & 255.6&181.4 & 327.5&224.4 & 243.7&161.9 & 248.5&178.6 \\
80 & 327.7&234.4 & 340.6&244.1 & 342.5&268.8 & 352.9&264.4 & 340.2&241.4 & 435.9&298.7 & 356.7&237.1 & 330.7&237.7 \\
85 & 436.2&312.0 & 453.4&324.9 & 455.9&357.8 & 469.7&351.9 & 452.7&321.3 & 527.4&361.4 & 474.8&315.6 & 440.2&316.4 \\
90 & 580.6&415.2 & 603.5&432.5 & 551.6&433.0 & 568.3&425.8 & 602.6&427.7 & 638.2&437.3 & 632.0&420.1 & 585.9&421.1 \\
95 & 772.8&552.7 & 730.2&523.3 & 667.5&523.9 & 756.4&566.7 & 729.1&517.5 & 772.2&529.1 & 764.7&508.3 & 708.9&509.6 \\
\hline
\end{tabular}\\
\vspace{3mm}
\end{table*}

\begin{table*}
\noindent
\centering
\caption{Paramerization of ART-XC PSF in a form of ellipse semi- major
  (a) and
  minor (b) axes with a corresponding ECF at angular distance 18'.}\label{tab:ecf:18}
\centering
\vspace{1mm}
 \begin{tabular}{|r|r|r|r|r|r|r|r|r|r|r|r|r|r|r|r|r|r|r|}
\hline
 &\multicolumn{16}{c|}{ART-XC mirror module}\\
 & \multicolumn{2}{c}{ARTM1} &  \multicolumn{2}{c}{ARTM2} 
 & \multicolumn{2}{c}{ARTM3} & \multicolumn{2}{c}{ARTM4} 
 & \multicolumn{2}{c}{ARTM5} & \multicolumn{2}{c}{ARTM6} 
 & \multicolumn{2}{c}{ARTM7} & \multicolumn{2}{c|}{ARTM8} \\
\hline
ECF & \multicolumn{1}{c|}{a} & \multicolumn{1}{c|}{b} &
\multicolumn{1}{c|}{a} & \multicolumn{1}{c|}{b} &
\multicolumn{1}{c|}{a} & \multicolumn{1}{c|}{b} &
\multicolumn{1}{c|}{a} & \multicolumn{1}{c|}{b} &
\multicolumn{1}{c|}{a} & \multicolumn{1}{c|}{b} &
\multicolumn{1}{c|}{a} & \multicolumn{1}{c|}{b} &
\multicolumn{1}{c|}{a} & \multicolumn{1}{c|}{b} & 
\multicolumn{1}{c|}{a} & \multicolumn{1}{c|}{b} \\
(\%) & \multicolumn{1}{c|}{('')}  & \multicolumn{1}{c|}{('')} &
\multicolumn{1}{c|}{('')} & \multicolumn{1}{c|}{('')} &
\multicolumn{1}{c|}{('')} & \multicolumn{1}{c|}{('')} &
\multicolumn{1}{c|}{('')} & \multicolumn{1}{c|}{('')} &
\multicolumn{1}{c|}{('')} & \multicolumn{1}{c|}{('')} &
\multicolumn{1}{c|}{('')}  & \multicolumn{1}{c|}{('')} &
\multicolumn{1}{c|}{('')} & \multicolumn{1}{c|}{('')} &
\multicolumn{1}{c|}{('')} & \multicolumn{1}{c|}{('')} \\
 \hline
5  &   9.5&  7.9 &   9.4&  7.6 &  10.2&  8.4 &   9.5&  7.5 &   9.7&  7.0 &   9.6&  8.5 &   8.4&  6.8 &   8.9&  7.0 \\
10 &  17.6& 10.5 &  17.4&  9.7 &  18.1& 11.3 &  17.9& 10.0 &  18.1&  9.4 &  17.3& 11.6 &  16.3&  8.7 &  15.9&  9.6 \\
15 &  25.7& 12.9 &  26.3& 12.1 &  27.2& 13.7 &  27.0& 12.4 &  27.2& 11.8 &  26.3& 14.0 &  25.1& 10.5 &  24.6& 11.4 \\
20 &  34.5& 15.8 &  35.1& 15.1 &  36.0& 16.5 &  35.9& 15.5 &  36.9& 15.2 &  36.1& 17.0 &  34.8& 13.6 &  33.7& 14.1 \\
25 &  42.7& 19.3 &  43.6& 18.8 &  44.2& 20.2 &  44.4& 19.4 &  46.2& 19.6 &  46.0& 21.0 &  44.1& 17.4 &  42.7& 17.8 \\
30 &  50.9& 23.4 &  52.1& 23.4 &  52.7& 24.4 &  53.0& 24.4 &  55.3& 24.3 &  56.2& 26.2 &  53.3& 22.1 &  51.6& 22.5 \\
35 &  59.4& 28.3 &  60.6& 28.6 &  61.0& 29.1 &  61.4& 29.5 &  64.7& 29.5 &  67.0& 32.2 &  62.2& 27.3 &  60.7& 27.8 \\
40 &  68.5& 33.6 &  69.7& 34.2 &  69.9& 34.3 &  70.7& 35.2 &  74.8& 35.3 &  78.6& 38.9 &  71.8& 32.8 &  70.6& 33.7 \\
45 &  78.3& 39.5 &  79.8& 40.4 &  80.0& 40.6 &  81.2& 41.9 &  86.0& 42.0 &  91.9& 46.8 &  82.4& 38.8 &  81.7& 40.4 \\
50 &  89.3& 46.0 &  91.1& 47.3 &  91.3& 48.4 &  92.5& 49.2 &  98.0& 49.4 & 107.8& 61.1 &  94.3& 45.7 &  93.7& 47.8 \\
55 & 102.2& 58.5 & 104.8& 60.8 & 106.1& 64.9 & 107.1& 65.1 & 113.5& 65.1 & 124.8& 78.1 & 108.5& 58.1 & 108.9& 63.5 \\
60 & 116.2& 74.3 & 119.4& 76.5 & 121.2& 82.1 & 121.9& 81.3 & 129.1& 81.0 & 166.2&104.0 & 123.9& 73.8 & 124.2& 79.3 \\
65 & 140.6& 89.8 & 144.4& 92.5 & 161.3&109.2 & 162.2&108.3 & 171.8&107.8 & 201.1&125.8 & 150.0& 89.3 & 165.3&105.5 \\
70 & 187.1&119.6 & 192.2&123.2 & 214.7&145.4 & 215.9&144.1 & 228.6&143.5 & 267.6&167.4 & 199.6&118.9 & 200.0&127.7 \\
75 & 274.0&175.1 & 281.4&180.3 & 285.8&193.5 & 287.3&191.8 & 304.3&191.0 & 356.2&222.8 & 265.7&158.3 & 292.9&187.0 \\
80 & 364.6&233.0 & 374.6&240.0 & 380.4&257.6 & 382.5&255.3 & 405.0&254.3 & 474.1&296.6 & 353.6&210.6 & 389.8&248.8 \\
85 & 441.2&282.0 & 453.3&290.4 & 460.2&311.7 & 462.8&308.9 & 490.1&307.7 & 573.7&358.9 & 470.7&280.4 & 471.7&301.1 \\
90 & 587.3&375.3 & 603.3&386.6 & 612.6&414.8 & 615.9&411.2 & 652.3&409.5 & 694.1&434.3 & 626.5&373.2 & 627.8&400.7 \\
95 & 781.6&499.5 & 730.0&467.8 & 741.2&501.9 & 745.3&497.5 & 789.3&495.5 & 763.6&477.7 & 758.0&451.5 & 759.6&484.9 \\
\hline
\end{tabular}\\
\vspace{3mm}
\end{table*}

\newpage

\bibliographystyle{spphys}       
\bibliography{references}   

\end{document}